\documentclass[aps,ams,amsmath,prl,twocolumn]{revtex4-1}
\usepackage{bm}
\usepackage{graphics}
\usepackage{graphicx}
\usepackage[dvipsnames]{xcolor}
\usepackage[colorlinks=true,linkcolor=blue,citecolor=BrickRed]{hyperref}
\begin{document}
	
	\title{Anomalous Transport Phenomena in $p_x+ip_y$ Superconductors}
	
	\author {Songci Li}
	
	\affiliation{Department of Physics, University of Washington, Seattle, WA 98195}
	
	\author{A. V. Andreev}
	
	\affiliation{Department of Physics, University of Washington, Seattle, WA 98195}
	
	\author {B. Z. Spivak}
	
	\affiliation{Department of Physics, University of Washington, Seattle, WA 98195}

	\begin{abstract}
		Spontaneous breaking of time-reversal symmetry in superconductors with the $p_x+ip_y$ symmetry of the order parameter allows for a class of effects which are analogous to the anomalous Hall effect in ferromagnets. These effects exist below the critical temperature, $T<T_{c}$.  We develop a kinetic theory of such effects. In particular, we consider anomalous Hall thermal conductivity,  the polar Kerr effect, the anomalous Hall effect, and the anomalous photo- and acousto-galvanic effects.
	\end{abstract}
	\date{\today}
	
	\maketitle

	\textit{Introduction:} One of the leading candidates for $p$-wave pairing in electronic systems
	is Sr$_{2}$RuO$_{4}$. There are numerous pieces of experimental
	evidence that the  superconducting state of Sr$_{2}$RuO$_{4}$ has odd parity,
	breaks time reversal symmetry and is spin
	triplet~\cite{nelson,kidwingira,Xia,Luke,rmp_pwave,maeno}. An order parameter
	consistent with these experiments is given by the chiral $p$-wave state
	\cite{sigrist} which is an analog of $^{3}$He-$A$. It has the form $\Delta_{\alpha\beta}( {\bf p}) \sim p_{x}\pm
	ip_{y}$ where $\Delta_{\alpha\beta} ({\bf p})$ is the Fourier transform of
	$\Delta_{\alpha\beta}({\bf r}-{\bf r}')$.
	However, the observation of power laws in specific
	heat \cite{Nishizaki} and NMR \cite{Ishida}, the absence of electric currents along
	edges \cite{Bjornsson}, and the absence of a split
	transition in the presence of an in-plane magnetic field \cite{Mao}
	are inconsistent with the theoretically expected
	properties of a simple chiral superconductor. Consideration of additional experimental manifestations of spontaneous breaking of time-reversal symmetry in $p_x+i p_y$ superconductors may clarify the nature of superconducting state in
	Sr$_{2}$RuO$_{4}$.

	Due to spontaneous breaking of time-reversal symmetry,  $p_x+ip_y$ superconductors must exhibit anomalous transport phenomena similar  to those which exist in metallic ferromagnets (see Refs.~\cite{AnomalousHallOng} and \cite{AnomalousHallNiu} for a review).
	In this article we develop a theory of several such effects in $p_x+ip_y$ superconductors: the anomalous Hall effect, polar Kerr effect  for microwave radiation, anomalous Hall thermal conductivity, and
	anomalous photo- and acousto-galvanic effects.

	It should be noted that $p$-wave superconductivity exists only in the clean regime, $l>\xi$, where electron transport may be described semiclassically. Generally, in the semiclassical regime there are three contributions to anomalous transport phenomena: skew scattering, side jumps, and the intrinsic contribution. The side jump contribution arises from the shift of the center of mass of electron wave packets during the scattering events, while the intrinsic contribution is related to the anomalous velocity due to Berry curvature. The magnitude of these contributions is independent of the mean free path. In contrast, the magnitude of the skew scattering contribution is proportional to the quasiparticle mean free path $l$. As a result, the skew scattering contribution exceeds the intrinsic and side jump contributions by a factor $\sim p_{F}l \gg 1$. Here $p_{F}$ is the Fermi momentum. Therefore in this article we will take into consideration only the skew scattering contribution.
	We focus on anomalous transport phenomena in the vicinity of the critical temperature, where quasiparticles play a major role.

	\textit{Kinetic scheme:} Transport theory in conventional time reversal invariant superconductors was developed long ago (see for example reviews Refs.~\cite{Aronov} and \cite{Aronov_bookchapter}). Below we generalize this approach to superconductors without time reversal symmetry, which exhibit anomalous transport phenomena. In the clean regime, $l\gg \xi$, and at sufficiently low frequencies, $\omega \ll |\Delta|$, where $|\Delta|$ is the modulus of the order parameter, the quasiparticle dynamics can be described by the Boltzmann kinetic equation for quasiparticles
	\begin{equation}\label{KinEq}
		\frac{\partial n_{{\bf p}}({\bf r},t)}{\partial t} +\frac{\partial \tilde{\epsilon}_{{\bf p}}}
		{\partial {\bf p}}\frac{\partial n_\mathbf{p}}{\partial {\bf r}}-\frac{\partial \tilde{\epsilon}_\mathbf{p}}{\partial {\bf r}}\frac{\partial n_\mathbf{p}}{\partial {\bf p}}=I_{st},
	\end{equation}
	where
	\begin{subequations}
		\label{epsilon}
		\begin{eqnarray}
			\tilde{\epsilon}_{\bf p}=\epsilon_{{\bf p}}+\mathbf{v}\cdot{\bf p}_{s}, \quad \epsilon_{{\bf p}}=\sqrt{\tilde{\xi}^{2}_{{\bf p}}+|\Delta|^{2}} , \\
			\tilde{\xi}_\mathbf{p}=\xi_{{\bf p}}+\Phi+ \frac{\mathbf{p}_s^2}{2m}, \quad \xi_{{\bf p}}=\frac{p^{2}}{2m}-\epsilon_{F}.
		\end{eqnarray}
	\end{subequations}
	In Eq.~(\ref{epsilon}) $m$ is the electron mass, while ${\bf p}_{s}$ and $\Phi$, are given by
	\begin{equation}
		\label{eq:p_s_Phi}
		\mathbf{p}_{s}=\frac{\hbar}{2} \boldsymbol{\nabla}\chi-\frac{e}{c}\mathbf{A}, \quad
		\Phi =\frac{\hbar}{2}\partial_{t}\chi+e\phi,
	\end{equation}
	where $\chi$ is the order parameter phase, and $\phi$ and $\mathbf{A}$ are the scalar and vector potentials.
	From Eq.~(\ref{eq:p_s_Phi}) one obtains the equation for the acceleration of the condensate,
	\begin{equation}\label{eq:condensate_acceleration}
		\partial_t \mathbf{p}_s=e \mathbf{E}+\boldsymbol{\nabla}\Phi.
	\end{equation}
	Equations~(\ref{KinEq})-(\ref{eq:p_s_Phi}) should be supplemented by the expression for the current density,
	\begin{eqnarray}
		{\bf j}&=&\frac{eN}{m}\, {\bf p}_{s}+e\int d\Gamma{\bf v} n_{\bf p}, \label{current}
	\end{eqnarray}
	and by the charge neutrality condition,
	\begin{equation}\label{Phi}
		\nu\Phi=\int d\Gamma\, \frac{\tilde{\xi}_\mathbf{p}}{\tilde\epsilon_\mathbf{p}}
		\, n_{\mathbf{p}},
	\end{equation}
	that relates the gauge invariant scalar potential and the odd in $\xi$ part of the quasiparticle distribution function, the self-consistency equation for the order parameter.
	Here $d\Gamma=V d^3p/(2\pi\hbar)^3$ ($V$ is the volume of the sample) and $\mathbf{v}=d\xi_\mathbf{p}/d\mathbf{p}$.

	We work in linear response to external perturbations, and neglect corrections to equilibrium value of  $|\Delta|$.
	The collision integral $I_{st}=I_{st}^{(el)}+I_{st}^{\epsilon}$ in Eq.~(\ref{KinEq}) describes both elastic and inelastic scattering. We will assume that $\tau_{\epsilon}\gg \tau$, where $\tau_{\epsilon}$ and $\tau$ are inelastic and elastic mean free time respectively. Therefore the main contribution to the aforementioned anomalous effects  comes from elastic scattering, which is described by the collision integral
	\begin{equation}\label{eq:I}
		I_{st}=\int \left(W_{\mathbf{p}\mathbf{p^{\prime}}}n_{\mathbf {p^{\prime}}}-W_{\mathbf{p^{\prime}}\mathbf{p}}n_{\mathbf {p}}\right)\delta\left(\tilde{\epsilon}_{\mathbf{p}}-\tilde{\epsilon}_{\mathbf {p^{\prime}}}\right)d\Gamma^{\prime}.
	\end{equation}
	
	Skew scattering of quasiparticles corresponds to the part of scattering probability in Eq.~(\ref{eq:I}) that is associated with breaking of time reversal symmetry,
	$\delta W_{\mathbf{p}\mathbf{p^{\prime}}} =W_{\mathbf{p}\mathbf{p^{\prime}}}- W_{-\mathbf{p^{\prime}}-\mathbf{p}} \neq 0$.
	Thus, all the aforementioned effects are proportional to $\delta W_{\mathbf{p}\mathbf{p^{\prime}}}$.
	Skew scattering arises beyond the lowest Born approximation for the scattering amplitude.  Below we consider point-like impurities.  In the normal state such impurities scatter electrons only in the $s$-wave channel and do not cause skew scattering.  Therefore in the superconducting state skew scattering of quasiparticles is entirely due to the breaking of time reversal symmetry by the $p_x+ip_y$ order parameter.   The elastic scattering probability for quasiparticles with energy $\epsilon$ can be characterized by $\xi\equiv \xi_\mathbf{p}$, $\xi^{\prime}\equiv \xi_{\mathbf{p}^{\prime}}=\pm \xi$ and the asimuthal angles $\varphi$, $\varphi^{\prime}$, which define the direction of $\mathbf{p}$ and $\mathbf{p}'$ in the $xy$-plane.   For simplicity, we assume cylindrical Fermi surface and obtain for the scattering probability (see appedix for details).
	\begin{equation}\label{eq:W_kk'}
		W_{\mathbf{p}\mathbf{p}^{\prime}}= W_0 +
		W_1\left[1-\cos\left(\varphi-\varphi^{\prime}+2\delta_\epsilon\right)\right].
	\end{equation}
	Here  $\delta_\epsilon$ is the energy-dependent scattering phase shift. It is related to the $s$-wave scattering phase shift $\delta_n$ in the normal state by
	\begin{equation}\label{eq:phase_shift}
		\delta_\epsilon=\arctan\frac{\delta_n\epsilon}{\sqrt{\epsilon^2-\Delta^2}}.
	\end{equation}
	We assume weak impurities, for which $\delta_n\approx \tan \delta_n=-\pi \nu V_0$ is small. Here $\nu$ is the density of states on the Fermi level and $V_0$ is the impurity pseudo-potential~\cite{LandauQM}.
	In this case  $W_0$ and $W_1$ are given by
	\begin{subequations}
		\label{eq:W_12}
		\begin{eqnarray}
			\label{eq:W_0}
			W_0(\xi,\xi^{\prime})&=& \frac{\zeta(\epsilon)}{2 \nu \tau}  \, \frac{(\xi+\xi^{\prime})^2}{2\epsilon^2},\\
			\label{eq:W_1}
			W_1(\xi,\xi^{\prime})&=&
			\frac{\zeta(\epsilon)}{2 \nu \tau}\,  \frac{\Delta^2}{\epsilon^2}
			.\label{Wkkprime}
		\end{eqnarray}
	\end{subequations}
	Here   $\tau^{-1}=2 \pi n_i\nu V^2_0$, with $n_i$ being the impurity density, is the elastic scattering rate in the normal state. The coefficient $\zeta(\epsilon)=(\epsilon^2-\Delta^2)/[\epsilon^2(1+\delta_n^2)-\Delta^2]$ represents the enhancement factor of the quasiparticle scattering cross-section over the normal state value.
	The first term in Eq.~(\ref{eq:W_kk'}),  $W_0$ given by Eq.~(\ref{eq:W_0}) has the same structure as in $s$-wave superconductors. It describes scattering only within the same (particle-like, $\xi>0$, or hole-like, $\xi<0$) branch and does not lead to branch imbalance relaxation.
	The second term, $W_1$ in Eq.~(\ref{eq:W_kk'}) is absent in $s$-wave superconductors. It leads leads to both skew scattering and scattering between branches of quasiparticle spectrum with different signs of $\xi$.  The skew scattering cross-section, described by the $\sin(\varphi-\varphi')\sin 2\delta_\epsilon$ term in Eq.~(\ref{eq:W_kk'}),  is energy-dependent. It follows from Eqs.~(\ref{eq:W_kk'}), (\ref{eq:phase_shift}), and (\ref{eq:W_1}) that it changes sign when impurity potential $V_{0}$ changes from repulsive to attractive.

	Below we consider linear response to several external perturbations and look for the quasiparticle distribution function in the form $n_{\mathbf{p}}=n^{(0)}+n^{(1)}_{\mathbf{p}}$, where $n^{(0)}$ is a locally equilibrium Fermi distribution, and $n^{(1)}_{\mathbf{p}}$ describes the deviation from equilibrium.  Noting that the collision integral (\ref{eq:I}) is nullified by an arbitrary function $n^{(0)}(\tilde\epsilon_{\mathbf{p}})$ we write the linearized Boltzmann equation in the form
	\begin{equation}\label{eq:linear_response_general}
		\mathcal{S}(\mathbf{p})=\int d\Gamma^{\prime}W_{\mathbf{p}\mathbf{p}^{\prime}}(n^{(1)}_{\mathbf{p}}-n^{(1)}_{\mathbf{p}^{\prime}})\delta(\epsilon_\mathbf{p} -\epsilon_\mathbf{p}^{\prime}),
	\end{equation}
	where the specific form of the source $\mathcal{S}(\mathbf{p})$ depends on the type of perturbation.

	\textit{Anomalous Hall thermal conductivity:}
	We first consider the Hall component of the thermal conductivity $\kappa_{xy}$ which describes the heat flux perpendicular to the direction ($x$-axis) of the temperature gradient. In this case the source
	term in Eq.~(\ref{eq:linear_response_general}) has the form,
	\begin{equation}\label{eq:S_thermal}
		\mathcal{S}(\mathbf{p})= -\frac{\xi}{T}\, \mathbf{v}\cdot\boldsymbol{\nabla}T\, \frac{\partial n^{(0)}}{\partial \epsilon}.
	\end{equation}
	The expression for the heat flux is
	\begin{eqnarray}
		\mathbf{j}^Q&=&\int d\Gamma\, \epsilon_\mathbf{p}\,\frac{\partial \epsilon_\mathbf{p}}{\partial \mathbf{p}} \, n^{(1)}_{\mathbf{p}}.\label{thermalflux}
	\end{eqnarray}
	Note that $\partial \epsilon_\mathbf{p}/\partial \mathbf{p}=\mathbf{v}\xi/\epsilon$ is the group velocity of the quasiparticles while $\mathbf{v}$ is the bare velocity as in a normal metal, $|\mathbf{v}|=v_F$. The solution of Eqs.~(\ref{eq:linear_response_general}), (\ref{eq:S_thermal})  has  the form
	\[
	n^{(1)}_{\mathbf{p}}=-\frac{\xi}{T}v_F\nabla T\frac{\partial n^{(0)}}{\partial \epsilon}\left[\alpha_s(\epsilon)\sin\varphi+\alpha_c(\epsilon)\cos\varphi\right].
	\]
	The Hall component of the thermal conductivity tensor, $\kappa_{xy}$, is determined by $\alpha_s(\epsilon)$ in the above expression, which is given by $\alpha_s(\epsilon)=a(\epsilon)/[b^2(\epsilon)+a^2(\epsilon)]$, with
	\begin{subequations}\label{eq:ab_def}
		\begin{eqnarray}
			a(\epsilon)&=& \frac{\zeta(\epsilon)}{2\tau} \, \frac{\Delta^2}{2\epsilon|\xi|}\,\sin2\delta_\epsilon,\\
			b(\epsilon)&=&\frac{|\xi|}{\epsilon\tau}+\frac{\zeta(\epsilon)}{2\tau}\frac{\Delta^2}{2\epsilon|\xi|} \left          (\cos2\delta_\epsilon+2\right).
		\end{eqnarray}
	\end{subequations}
	For weak impurities, $|\delta_n| \ll 1$, we obtain, close to $T_c$
	\begin{equation}
		\kappa_{xy}= 3\kappa\left(\frac{\Delta}{\pi T}\right)^2 \delta_n, \label{kappaxy}
	\end{equation}
	where $\kappa=\pi^2 \nu T D/3$ (with $D=v_F^2\tau/2$ being the diffusion constant) is the normal state thermal conductivity.

	\textit{Polar Kerr effect:}
	Next we consider a linearly polarized electromagnetic wave at normal incidence to the $xy$ surface of $p_x+ip_y$ superconductor. The reflected wave is elliptically polarized with the major axis rotated with respect to the incident one by the polar Kerr angle \cite{longrangeorder}
	\begin{equation}
		\theta_k=\frac{(1-n^2+\kappa^2)\Delta\kappa+2n\kappa\Delta n}{(1-n^2+\kappa^2)^2+(2n\kappa)^2},
	\end{equation}
	where $n$ and $\kappa$ are, respectively, the real and imaginary part of the refraction index and
	\begin{eqnarray}
		\Delta n + i \Delta \kappa&=& -\frac{4\pi}{\omega} \frac{(n - i \kappa)\sigma_{xy}}{n^2+\kappa^2} ,
	\end{eqnarray}
	where $\sigma_{xy}$ is the complex ac conductivity.

	In this case  the electric field is uniform in the direction parallel to the surface of the sample, $\Phi=0$, and the value of $\mathbf{p}_{s}(t)$ is determined by Eq.~(\ref{eq:condensate_acceleration}). The diagonal part of the conductivity is given by~\cite{Aronov}
	\begin{equation}\label{eq:sigma_xx}
		\sigma_{xx} \approx\sigma_D + \frac{i N_s(T)}{\omega},
	\end{equation}
	where $N_s(T)$ is the temperature dependent superfluid density and $\sigma_D=e^2 \nu D $ is the Drude conductivity.
	In contrast to the thermal conductivity consideration, in the present case $n^{(0)}=1/(\exp[\epsilon_\mathbf{p}+\mathbf{v}\cdot{\bf p}_{s}(t)]/T+1)$ gives a nonvanishing contribution to the current response (via $N_s(T)$) because the superfluid momentum depends on the electric field. The Kerr angle $\theta$ is determined by the value of the Hall component of conductivity $\sigma_{xy}$.

	To find $\sigma_{xy}$ we seek the solution of Eq.~(\ref{KinEq}) in the form
	$n=n^{(0)}(\epsilon_{\mathbf{p}}/T)+n^{(1)}_{\mathbf{p}}$.
	The source in Eq.~(\ref{eq:linear_response_general}) becomes
	\begin{eqnarray}
		\mathcal{S}(\mathbf{p})&=&-i\omega n_{\mathbf{p}}^{(1)}-e\mathbf{v}\cdot\mathbf{E}\frac{\partial n^{(0)}}{\partial \epsilon}, \label{nk1}
	\end{eqnarray}
	where the external electric field $\mathbf{E}$ is along $x$-direction. The nonequilibrium distribution $n^{(1)}_{\mathbf{p}}$ has the form
	\begin{equation}\label{ansatzn1}
		n^{(1)}_{\mathbf{p}}=-eEv_F\frac{\partial n^{(0)}}{\partial \epsilon}\left[\beta_s(\epsilon)\sin\varphi+\beta_c(\epsilon)\cos\varphi\right].
	\end{equation}
	The Hall conductivity depends only on the function $\beta_s(\epsilon)$, which is given by
	$\beta_s(\epsilon)=a(\epsilon)/\{[b(\epsilon)-i \omega]^2+a^2(\epsilon)\}$, with $a(\epsilon)$  and $b(\epsilon)$ being defined in Eq.~(\ref{eq:ab_def}).
	Substituting Eq.~(\ref{ansatzn1}) into Eq.~(\ref{current}) we obtain the Hall conductivity in the weak impurity limit, $|\pi \nu V_0| \ll 1$, in the form
	\begin{eqnarray}
		\sigma_{xy}(\omega)&=&\sigma_D\frac{\Delta}{2T}\delta_n\int_{0}^{\infty}\frac{dx}{\cosh^2(\sqrt{x^2+1}\Delta/2T)}\nonumber\\
		&&\times\frac{x^2 +1}{(-i\omega\tau x\sqrt{x^2+1} + x^2 +3/4)^2}.\label{eq:sigmaxyintegralanytemperature}
	\end{eqnarray}
	where $x=|\xi|/\Delta$. At temperature close to $T_c$  and at
	low frequencies, $\omega \tau \ll 1$, this expression yields
	\begin{equation}\label{eq:sigma_xy_dc}
		\sigma_{xy}=\frac{7\pi}{12 \sqrt{3}}\, \delta_n \,\frac{\Delta}{T}\, \sigma_D.
	\end{equation}
	This result was derived assuming $p_x+ ip_y$ symmetry of the order parameter. In the $p_x-ip_y$ state the Hall conductivity $\sigma_{xy}$ has opposite sign. It also changes sign if the impurity potential $V_0$ changes from repulsive, $\delta_n<0$, to attractive, $\delta_n>0$, in agreement with Ref.~\cite{Lutchyn}. 
	Note that our result for the low frequency Hall conductivity, Eq.~(\ref{eq:sigma_xy_dc}), is proportional to the elastic mean free time $\tau$  and to the density of quasiparticles.
	
	There is another contribution to $\sigma_{xy}$ associated with the existence of the transverse component of the superfluid velocity $v_{y}\sim \dot{p}_{x}$, which is proportional to the condensate acceleration in the $x$-direction. It may not be obtained within the present formalism that is based on the Boltzmann kinetic equation for the quasiparticles. At $T\sim T_{c}$ this contribution is smaller than the quasiparticle contribution, Eq.~(\ref{eq:sigma_xy_dc}). However at $T\ll T_{c}$  when the quasiparticle contribution becomes exponentially small in Eq.~(\ref{eq:sigmaxyintegralanytemperature}) it becomes the dominant contribution.
	The requirement for this contribution to exist is violation of Galilean invariance in the system.
	Thus it should exist in any crystalline superconductors with $p_x+ip_y$ symmetry \cite{Kallin1,Kallin}.
	It can also be caused by electron-impurity scattering. In this case this contribution is inversely proportional to the electron mean free time~\cite{Lutchyn}.

\begin{figure}
		\includegraphics[width=8.0cm]{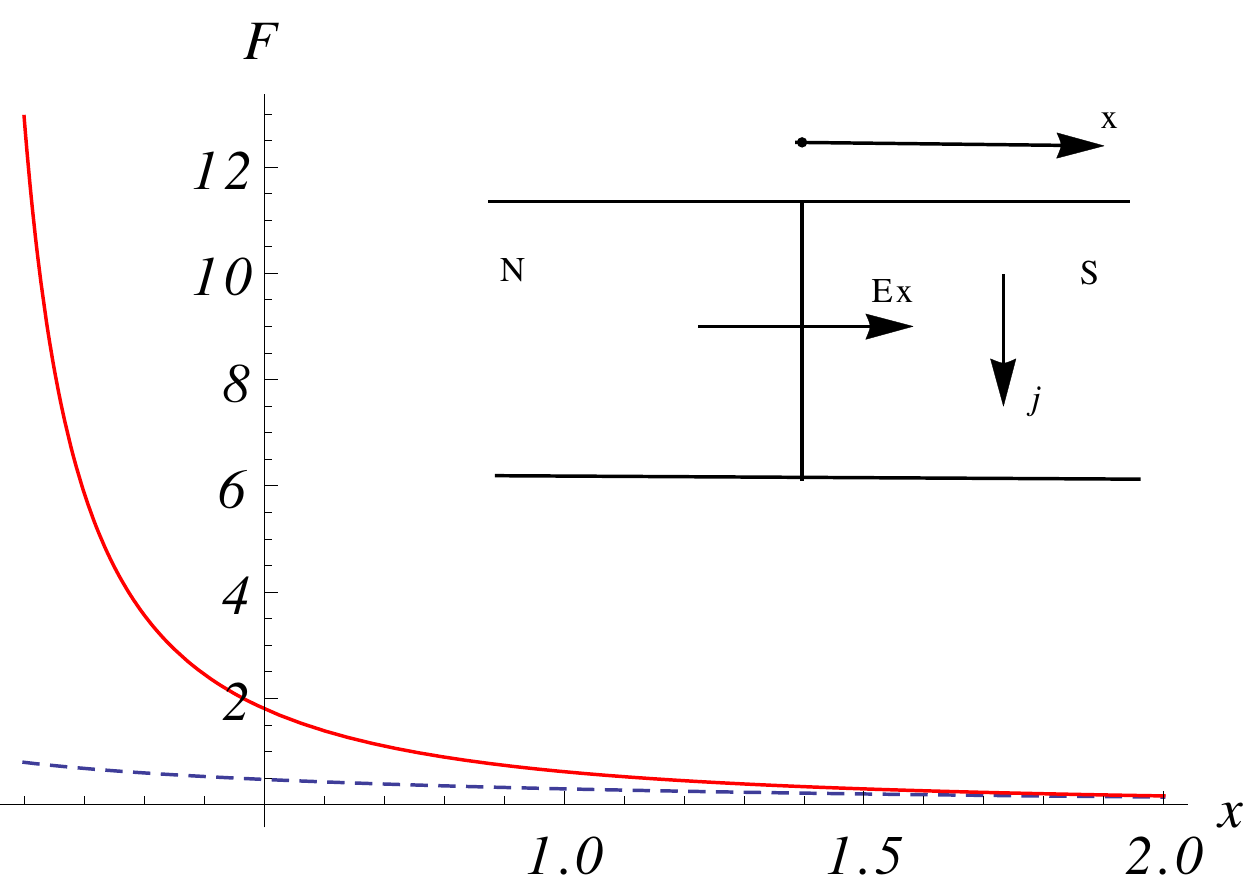}\\
		\caption{Plot of the functions $F_{-2}(x)$ (solid line) and $F_0(x)$ (dashed line) in Eq.~(\ref{eq:F_def}). The inset shows a schematic setup of the normal current injection experiment. Electric current is injected into the superconductor $S$ from the normal metal $N$ along along the $x$-axis. Skew scattering of quasiparticles generates an anomalous Hall current in the $y$ direction.}\label{fig:F_plot}
	\end{figure}

	\textit{Hall effect for normal current injection:}
	Let us now consider a normal metal/$p_x+ip_y$-superconductor junction, through which a steady current is flowing. At  $T \ll \Delta$ this situation was considered in Ref.~\cite{Keles}. In this regime conversion of normal current to supercurrent is mediated by multiple Andreev reflections. Here we work near the critical temperature and consider a setup, in which the normal current is  injected into the superconductor in the $x$- direction, as shown in the inset in Fig.~\ref{fig:F_plot}. In this case the conversion of quasiparticle current to the supercurrent occurs in the superconductor. Just as in the case of $s$-wave superconductor, near $T_c$, the electric field penetrates into superconductor to a large distance $L_{Q}\gg l$, which is determined by the relaxation of imbalance between the populations of quasiparticles in electron-like, $\xi>0$, and hole-like, $\xi<0$ branches of the spectrum (see for example Ref.~\cite{Aronov} and references therein). The new feature of normal current injection that appears in $p_x+ip_y$ superconductors is that skew scattering of quasiparticles generates nonequilibrium current that is perpendicular to the electric field. Another aspect is that, in contrast to s-wave superconductors, impurity scattering leads to branch imbalance relaxation even if the magnitude of the order parameter $|\Delta|$ is isotropic in the Fermi surface. Below we assume that the inelastic scattering rate is smaller than $1/\tau$ and thus impurity scattering gives the dominant contribution to branch imbalance relaxation.
	
	In linear response we write the quasiparticle distribution function in the superconductor in the form $n^{(0)}(\epsilon_{\mathbf{p}}/T)+ n^{(1)}_\mathbf{p}$. This yields the source term in Eq.~(\ref{eq:linear_response_general})
	\begin{equation}\label{eq:S_imbalance_near}
		\mathcal{S}(\mathbf{p}) =\frac{\xi}{\epsilon}\,\mathbf{v}\cdot\frac{\partial n_{\mathbf{p}}^{(1)}}{\partial \mathbf{r}} .
	\end{equation}
	At length scales in excess of the mean free path we may employ the diffusive approximation. With the aid of Eq.~(\ref{current})
	the Hall current $j_y$ can be expressed in the form
	\begin{equation}\label{eq:current_grad_n}
		j_y(x)=-4 e\nu D \, \delta_n\int d\xi \frac{\Delta^2}{\xi^2}\partial_x\bar{n}_{a}(\xi,x),
	\end{equation}
	where $\bar{n}_{a}(\xi,x)$ is the  antisymmetric in $\xi$ part of the distribution function averaged over the momentum directions. The latter satisfies the diffusion equation with relaxation
	\begin{equation}\label{eq:diffusionnabar}
		D\frac{\partial^2}{\partial x^2}\bar{n}_{a}(\xi,x)=\frac{1}{\tau_Q(\xi)}\bar{n}_{a}(\xi,x),\\
	\end{equation}
	with energy dependent relaxation rate $\tau_Q^{-1}(\xi)=\tau^{-1}\Delta^2(\xi^2+2\Delta^2)/\xi^4$.
	The solution of Eq.~(\ref{eq:diffusionnabar}) is
	\begin{equation}\label{eq:nabar}
		\bar{n}_{a}(\xi,x)=\bar{n}_{a}(\xi,0)\exp\left[-x/L_Q(\xi)\right],
	\end{equation}
	where $L_Q(\xi)=\sqrt{D\tau_Q(\xi)}$ is the energy-dependent branch imbalance relaxation length.
	
	We work in the vicinity of the critical temperature $T_c$, where for typical thermal quasiparticles  $\xi\sim T$, ($|\xi|\gg\Delta$), the relaxation lengths are long,  $L_Q(\xi)=l|\xi|/\Delta \gg l$. These quasiparticles diffuse into the bulk of the superconductor and contribute to the gauge-invariant potential $\Phi$ given by Eq.~(\ref{Phi}). The boundary value of the nonequilibrium quasiparticle population, $\bar{n}_{a}(\xi,0)$ in Eq.~(\ref{eq:nabar}) is obtained by matching the solution of Eq.~(\ref{eq:diffusionnabar}) with the solution of diffusion equation with energy relaxation for the electrons in the normal metal. The result depends on both the inelastic mean free path in the normal metal, $l_e$, and the branch imbalance relaxation length $L_Q$ in the superconductor.  For $l_e \gg L_Q$ the boundary condition is 
	\begin{equation}\label{barnaxix}
		\bar{n}_{a}(\xi,0)= \mathrm{sign}(\xi)\frac{eE_x(0)L_Q(\xi)}{4T\cosh^2(\xi/2T)},
	\end{equation}
	where $E_x(0)$ is the electric field in the normal metal generating the steady current. Here we used the fact that  in the stationary case ${\bf E}=-\boldsymbol{\nabla} \Phi/e$, which follows  from Eq.~(\ref{eq:condensate_acceleration}). Using this relation and
	substituting Eqs.~(\ref{barnaxix}), (\ref{eq:nabar}) into Eqs.~(\ref{Phi}) and (\ref{eq:current_grad_n}) we obtain the spatial distributions of the electric field $E_x (x)$ and the Hall current $j_y(x)$ in the superconductor,
	\begin{eqnarray}\label{Phix}
		E_x (x)&=& E_x (0) F_0\left(\frac{x}{\langle L_Q \rangle}\right),\\
		j_y(x)&=&\sigma_D E_x (0)\delta_n\left(\frac{\Delta}{T}\right)^2F_{-2}\left(\frac{x}{\langle L_Q \rangle}\right), \label{eq:j_y_x}
	\end{eqnarray}
	where $\langle L_Q \rangle=2\ln2(Tl/\Delta)$ and the functions $F_{n}$, are defined as
	\begin{equation}\label{eq:F_def}
		F_n(x)=\int_{0}^{\infty}dy\frac{y^n}{\cosh^2(y)}\exp\left(-\ln2\, \frac{x}{y}\right),
	\end{equation}
	and are plotted in Fig.~\ref{fig:F_plot}. The spatial distributions of the Hall current $j_y(x)$ and the electric field $E_x(x)$ are drastically different, and cannot be related by a local Hall conductivity $\sigma_{xy}$. At relatively short distances, $l\ll x\ll \langle L_Q\rangle $, we see from Eq.~(\ref{eq:j_y_x}) that $j_y(x)\propto \sigma_D E_x(0) \delta_n (\Delta/T)^2 \langle L_Q\rangle/x $, so that the Hall current is
	\[
	I_y=\int dx j_y(x)\approx \sigma_D E_x (0)\delta_n\left(\frac{\Delta}{T}\right)^2\langle L_Q \rangle\ln\frac{\langle L_Q \rangle}{l}.
	\]
	
	\textit{Anomalous photo- and acousto-galvanic effects:}
	When an electromagnetic or an acoustic wave propagates through a conductor it generates an anisotropic in momentum ${\bf p}$  distribution function. The induced current density is proportional to the rate of the the momentum transfer from the wave to the electron system~\cite{Aronov_bookchapter},
	\[
	J_{x}=I\alpha_{xx}.
	\]
	Here $I$ is the rate of momentum density transfer due to the wave adsorption, and  $x$ is the direction of the wave propagation.
	In $p_x+ip_y$ superconductors an anomalous current in the $y$ direction is generated. Considerations similar to those leading to Eq.~(\ref{eq:sigma_xy_dc}) near $T_{c}$ yield
	\[
	\alpha_{xy}\sim \alpha_{xx} \frac{\Delta}{T}\,\delta_n.
	\]

	Finally, we note that all anomalous transport phenomena discussed above are driven by the underlying symmetry of the superconducting state. Therefore they should exist in any superconductor whose order parameter breaks time reversal symmetry, see for example~\cite{Sauls,Strand,Strand1,Schemm,Schemm1}. Although our consideration focused on $p_x+ip_y$ materials we believe our approach is applicable to other superconductors with broken time-reversal invariance.
	
	\acknowledgments   S. L. and A. A. were supported by the U. S. Department of Energy Office of Science, Basic Energy Sciences under award number
	DE-FG02-07ER46452. B.S. thanks the International Institute of
	Physics (Natal, Brazil) for hospitality.
	
\appendix*
	\section{Appendix: Derivation of the scattering probability}
	\setcounter{equation}{0}    
	\renewcommand{\theequation}{A.\arabic{equation}}
	In this appendix we obtain the scattering probability $W_{\mathbf{p}\mathbf{p}^{\prime}} $ in Eq.~(\ref{eq:W_kk'}) that gives rise to anomalous transport properties. For simplicity we consider point-like impurities whose scattering matrix elements are independent of momentum, $V_{\mathbf{p}\mathbf{p}^{\prime}}=V_0$. Skew scattering appears beyond the lowest Born approximation as a result of a particular structure of the order parameter in $p_x+ip_y$ superconductors. We note that even weak impurities, $|\nu V_0| \ll 1$,
	lead to the existence of bound states in $p_x+ip_y$ superconductors with binding energies $E_b\sim |\Delta| |\nu V_0|^2$~\cite{Okuno_Sigrist}. Therefore low energy quasiparticles undergo resonant scattering. We show below that this also results in energy-dependent skew scattering.
	
	The scattering properties of a single impurity are described by the $T$-matrix. In a superconductor, for each initial and final momenta $\mathbf{p}^{\prime}$ and $\mathbf{p}$ the $T$-matrix acquires an additional $2\times 2$ structure in the Nambu-Gorkov space.
	We denote these $2\times 2$ matrices by $\hat{T}_{\mathbf{p}\mathbf{p}^{\prime}}$. The quasiparticle scattering amplitude is given by the on-shell matrix element  $\hat{T}_{\mathbf{p}\mathbf{p}^{\prime}}\equiv\langle \mathbf{p}|\hat{T}|\mathbf{p}^{\prime}\rangle $, with $\left| \mathbf{p} \right\rangle$ being the two component quasiparticle wave function.
	Using the Fermi Golden rule the scattering probability can be expressed as
	\begin{equation}\label{scatproba}
	W_{\mathbf{p}\mathbf{p}^{\prime}}=2\pi n_i|\hat{T}_{\mathbf{p}\mathbf{p}^{\prime}}|^2
	\end{equation}
	where $n_i$ is the impurity concentration. The $T$-matrix obeys the Lipmann-Schwinger equation
	\begin{equation}\label{Tmatrix}
	\hat{T}_{\mathbf{p}\mathbf{p}^{\prime}}=\hat{V}_{\mathbf{p}\mathbf{p}^{\prime}}+
	\sum_{\mathbf{p}^{\prime\prime}}\hat{V}_{\mathbf{p}\mathbf{p}^{\prime\prime}}
	\hat{G}_{0}(\mathbf{p}^{\prime\prime})\hat{T}_{\mathbf{p}^{\prime\prime}\mathbf{{p}^{\prime}}},
	\end{equation}
	For point-like impurities, $\hat{V}_{\mathbf{p}\mathbf{p}^{\prime}}=V_0\hat{\tau}_3$ (with $\hat{\mathbf{\tau}}_i$ being the Pauli matrices in the Nambu-Gorkov space),   the $T$-matrix depends only on the energy $E$. Thus Eq.~(\ref{Tmatrix}) simplifies to
	\begin{equation}\label{TmatrixE}
	\hat{T}(E)=V_0\hat{\tau}_3+V_0\hat{\tau}_3\sum_{\mathbf{p}}\hat{G}_{0}(\mathbf{p})\hat{T}(E)
	\end{equation}
	
	The order parameter of $p_x+ip_y$ superconductor can be expressed as $\Delta(p_x+ip_y)/p_F=\Delta \exp(i\varphi_{\mathbf{p}})$ where $\varphi_{\mathbf{p}}$ is the momentum-dependent phase such that $\cos\varphi_{\mathbf{p}}=p_x/p_F$ and $\sin\varphi_{\mathbf{p}}=p_y/p_F$. This leads to the BCS Hamiltonian
	\begin{equation}
	\hat{H}(\mathbf{p})=\left(
	\begin{array}{cc}
	\xi_{\mathbf{p}} & \Delta e^{i\varphi_{\mathbf{p}}}\\
	\Delta e^{-i\varphi_{\mathbf{p}}} & -\xi_{\mathbf{p}} \\
	\end{array}
	\right),
	\end{equation}
	and the Green function $\hat{G}_{0}(\mathbf{p})\equiv\left(E-\hat{H}(\mathbf{p})\right)^{-1}$ ,
	\begin{eqnarray}
	\hat{G}_{0}(\mathbf{p})&=&\frac{1}{E^2-\epsilon_{\mathbf{p}}^2}
	\left(\begin{array}{cc}
	E+\xi_{\mathbf{p}} & \Delta e^{i\varphi_{\mathbf{p}}} \\
	\Delta e^{-i\varphi_{\mathbf{p}}} & E-\xi_{\mathbf{p}} \\
	\end{array}\right).\label{green}
	\end{eqnarray}
	Replacing $\sum_{\mathbf{p}}$ with $V\int d^3p/(2\pi\hbar)^3\rightarrow\nu\int d\xi_{\mathbf{p}}\int d\varphi_{\mathbf{p}}/2\pi$, where $V$ is the volume of the sample and $\nu$ is the density of state at Fermi surface, and using Eq.~(\ref{TmatrixE}) and Eq.~(\ref{green}), we obtain for the $T$-matrix,
	\begin{equation*}
	\hat{T}(E)=\left(
	\begin{array}{cc}
	\frac{V_0}{1+\pi\nu V_0E/\sqrt{\Delta^2-E^2}} & 0 \\
	0 & \frac{-V_0}{1-\pi\nu V_0E/\sqrt{\Delta^2-E^2}} \\
	\end{array}
	\right)
	\end{equation*}
	The pole of the $T$-matrix describes the subgap bound state~\cite{Okuno_Sigrist} with the binding energy $E_b= \Delta/\sqrt{1+(\pi\nu V_0)^2}$.  For the energy in the continuum, $|E|> \Delta$, we obtain the $T$-matrix in the form
	\begin{equation}
	\hat{T}(E)=\left(
	\begin{array}{cc}
	f(E)e^{i\delta_E} & 0 \\
	0 & -f(E)e^{-i\delta_E} \\
	\end{array}
	\right)
	\end{equation}
	where  $\delta_E$ is given by
	Eq.~(\ref{eq:phase_shift}) and
	\begin{equation}\label{eq:f}
	f(E)=\frac{V_0}{\sqrt{1+(\pi\nu V_0)^2E^2/(E^2-\Delta^2)}}.
	\end{equation}

	The quasiparticle scattering amplitude is given by the on-shell matrix elements between two quasiparticle states with energy
	$\epsilon=\sqrt{\Delta^2+\xi^2_{\mathbf{p}}}=\sqrt{\Delta^2+\xi^2_{\mathbf{p}^{\prime}}} $,
	\begin{equation}\label{eq:T_amplitude}
	\hat{T}_{\mathbf{p}\mathbf{p}^{\prime}}=\langle\mathbf{p}|\hat{T}(\epsilon)|\mathbf{p}^{\prime}\rangle.
	\end{equation}
	In a $p_x +i p_y$ superconductor the Bogoliubov amplitudes carry a momentum-dependent phase $\varphi_\mathbf{p}$,
	\begin{equation}\label{eq:spinor}
	\langle \mathbf{p} | =(u_{\mathbf{p}} e^{-i \varphi_\mathbf{p}}, v_{\mathbf{p}} ),
	\end{equation}
	where $u_{\mathbf{p}}$ and $v_{\mathbf{p}}$ have the standard BCS form and are independent of the momentum direction,
	\begin{equation}\label{eq:uv}
	u_{\mathbf{p}}=\sqrt{\left(1+ \xi_{\mathbf{p}}/\epsilon_{\mathbf{p}}\right)/2}, \quad v_{\mathbf{p}}=\sqrt{\left(1- \xi_{\mathbf{p}}/\epsilon_{\mathbf{p}}\right)/2}
	\end{equation}
	Substituting Eqs.~(\ref{eq:spinor}) and (\ref{eq:uv}) into (\ref{eq:T_amplitude}) we get
	\begin{eqnarray}
	\left| \hat{T}_{\mathbf{p}\mathbf{p}^{\prime}}\right|^2
	&=&f^2(\epsilon)\Big\{\left(u_{\mathbf{p}}u_{\mathbf{p}^{\prime}}-v_{\mathbf{p}}v_{\mathbf{p}^{\prime}}\right)^2+2u_{\mathbf{p}}u_{\mathbf{p}^{\prime}}v_{\mathbf{p}}v_{\mathbf{p}^{\prime}}
	\nonumber \\
	&&\times\left[1-\cos(2\delta_\epsilon+\varphi_{\mathbf{p}}-\varphi_{\mathbf{p}^{\prime}})\right]\Big\}\label{Tkkprimesquare}\\
	&=&\frac{1}{2}f^2(\epsilon)\left(1+\frac{\xi_{\mathbf{p}}\xi_{\mathbf{p}^{\prime}}-\Delta^2}{\epsilon_{\mathbf{p}}^2}\right)+ \nonumber \\
	&&f^2(\epsilon)\frac{\Delta^2}{2\epsilon_{\mathbf{p}}^2}\Big[1-\cos\left(2\delta_\epsilon+\varphi_{\mathbf{p}}-
	\varphi_{\mathbf{p}^{\prime}}\right)\Big]
	\end{eqnarray}
	Substituting this result  into Eq.~(\ref{scatproba}), and recalling that for weak impurities, $|\pi \nu V_0|\ll 1$, the scattering rate in the normal state is $\tau^{-1}=2 \pi n_i\nu V^2_0$
	we obtain the scattering probability in the form of Eq.~(\ref{eq:W_kk'}) and Eq.~(\ref{eq:W_12}).

\end{document}